# Physical property characterization of Fe-tube encapsulated and vacuum annealed bulk $MgB_2$

V.P.S. Awana[1,2,$], Rajeev Rawat[3], Anurag Gupta[1], M. Isobe[2], K.P. Singh[1], Arpita Vajpayee[1], H. Kishan[1], E. Takayama-Muromachi[2] and A.V. Narlikar[3]

[1]*National Physical Laboratory, Dr. K.S. Krishnan Marg, New Delhi-110012, India*

[2]*Superconducting Materials Center, NIMS, 1-1 Namiki, Tsukuba, Ibaraki, 305-0044, Japan*

[3]*UGC-DAE Consortium for Scientific Research, University Campus, Khandwa Road, Indore-452017, India*

We report phase formation, and detailed study of magnetization and resistivity under magnetic field of $MgB_2$ polycrystalline bulk samples prepared by Fe-tube encapsulated and vacuum ($10^{-5}$ torr) annealed (750 $^0$C) route. Zero-field-cooled magnetic susceptibility ($\chi^{ZFC}$) measurements exhibited sharp transition to superconducting state with a sizeable diamagnetic signal at 39 K ($T_c$). The measured magnetization loops of the samples, despite the presence of flux jumps, exhibited a stable current density ($J_c$) of around 2.4 x $10^5$ A/cm$^2$ in up to 2 T (Tesla) field and at temperatures ($T$) up to 10 K. The upper critical field is estimated from resistivity measurements in various fields and shows a typical value of 8 T at 21 K. Further, $\chi^{FC}$ measurements at an applied field of 0.1 T reveal paramagnetic Meissner effect (*PME*) that is briefly discussed.

$: Corresponding Author: awana@mail.nplindia.ernet.in

## INTRODUCTION

Soon after the discovery of superconductivity in $MgB_2$ with the transition temperature of ~ 40 K[1], the focus changed within a year from basic to applied research in terms of improving the critical current density ($J_c$) of the compound for practical purposes[2-5]. Clean $MgB_2$ compounds having high residual resistivity ratio (RRR), defined as $R_{300K}/R_{40K}$ exhibited low critical current density ($J_c$) values. This limited the scope of this compound for any viable practical use. However soon it was realized that induction of disorder could result in record high $J_c$ values with only a marginal reduction in $T_c$[3-6].

So far, a number of techniques have been developed to increase the $J_c$ of $MgB_2$,: for example, (a) various nano-particle doping viz. nano-SiC[3,4], carbon-nanotubes[5] and nano-diamond[6], and (b) following various heating schedules viz. liquid assisted sintering[7] and combustion[8]. Similar novel methods had earlier been tried in case of high $T_c$ cuprate superconductors, but were not as successful as in the case of $MgB_2$. The ultimate aim had been to create optimally distributed pinning centers in a superconductor to inhibit the motion of vortices and thereby achieve a high $J_c$. By now, via the various techniques of doping/substitutions or heating methodology, the $J_c$ values in excess of $10^6$ A/cm$^2$ have already been achieved for $MgB_2$[3-8]. There is a race currently in progress to produce high quality $MgB_2$ powder for making wires and tapes[3-8] to carry higher transport current densities. In this regards, we report here dc susceptibility, magnetization and magnetoresistivity characterization of bulk polycrystalline $MgB_2$ compounds produced by Fe-tube encapsulated and vacuum ($10^{-5}$ torr) annealed (750 $^0$C) route.

## EXPERIMENTAL

Our $MgB_2$ samples were synthesized by encapsulation of well mixed and palletized high quality (above 3 N purity) Mg and B powders with some added Mg turnings in a soft Fe-tube and its subsequent heating to 750 $^0$C for two and half hours in an evacuated (10$^{-5}$ Torr) quartz tube and quenching to liquid-nitrogen temperature. Details of synthesis procedure are given elsewhere[9,10]. In our case the resultant sample was a bulk polycrystalline compound. The x-ray diffraction pattern of the compound was recorded with a diffractometer using CuK$_\alpha$ radiation. Resistivity measurements are carried out by four-probe technique under applied field of up to 8 Tesla. Magnetization measurements were carried out with a Quantum-Design SQUID magnetometer *(MPMS-XL)*.

Fig. 1 depicts the x-ray diffraction pattern of the currently studied Fe-tube encapsulated and vacuum (10$^{-5}$ torr) annealed MgB$_2$ polycrystalline bulk sample. The compound has hexagonal Bravais lattice with lattice parameters of a = 3.086 Å, and c = 3.524 Å. It is clear from this figure that the compound is nearly single phase, with small un-reacted line at around 2θ ≅ 36° and 63°. The one at 2θ ≅ 36° is due to a Mg metal[11] and other at 63° is due to presence of MgO, as suggested earlier by various authors[4,11].

Resistivity versus temperature $\rho$ (T) plots of our MgB$_2$ compound, are presented in Fig. 2 and Fig. 3. Namely $\rho$(T) plots are shown in Fig. 2 and Fig.3 depicts the d$\rho$/dT(T) plots in various applied fields of up to 8 Tesla. The extended part of $\rho$(T) plots is further shown for transition region in inset of Fig. 2. Single d$\rho$/dT(T) peaks for all measurements under various applied fields indicate towards good quality of the present sample. The superconducting transition temperatures ($T_c$) are defined by the peak temperature of d$\rho$/dT(T) plots, which in turn determine the upper critical fields ($H_{c2}$) at those temperatures. The plot of $H_{c2}$(T) being estimated with such a reasoning is shown in inset of Fig. 3. The upper critical field shows a typical value of 8 T at 21 K.

Fig.4 depicts the dc susceptibility ($\chi$) versus temperature plots for the MgB$_2$ sample in an applied field of 5 Oe, in both zero-field-cooled (ZFC) and field-cooled (FC) situations. It is evident from this figure that the present MgB$_2$ undergoes a sharp superconducting transition (diamagnetic) at around 39 K within less than 2 K temperature interval, without any rounding before saturation down to 5 K. In fact, the diamagnetic signal remains more or less constant below 36 K down to 5 K. The $\chi^{FC}$ signal around $T_c$ is weak and indicates strong pinning in the samples. Interestingly, the $\chi^{FC}$ (*T*) plot in higher applied field of 0.1 T, as shown in the inset of Fig.4, exhibits a paramagnetic transition around 37 K. This effect has been called Paramagnetic Meissner Effect (*PME*). The thermal cycling of $\chi^{FC}$ (*T*) in terms of *FCC* (*FC*-cooling) and *FCW* (*FC*-warming) exhibits hysteresis[12], as evident in inset of Fig. 4, where $\chi^{FC}$ (*T*) measurements are carried out in 0.1 Tesla field in both *FCC* and *FCW* situations. The details of *PME* phenomenon, shall be a subject of separate article. However, the presence of *PME* does imply that the sample must be containing intrinsic SIS (Superconductor-Insulator-Superconductor) or SNS (Superconductor-Normal-Superconductor) junctions[12,13]. It is argued earlier that pi-junctions of SIS/SNS in favorable conditions can give rise to PME. The presence of micron or smaller normal/insulating impurities, for instance Mg/MgO (as seen in XRD, Fig.1) could act as possible SIS/SNS junction to favor the observed *PME* in the samples. Interestingly, these nano-metric impurities might very well act as effective pinning centers to give high $J_C$ in the samples (see below).

In Fig. 5 we show the magnetic hysteresis *M(H)* loops for our 750 $^0$C annealed MgB$_2$ compound at 5, 10, 20, 25 and 30 K with applied fields (*H*) of up to ± 7 Tesla. As is evidenced from this figure, there are some interesting points to note that (i) despite the flux jumps, the magnetization of our samples is nearly invariant, for up to $T \leq 20$ K, in the field range of 1 T > *H* > -1 T and (ii) the irreversibility field ($H_M^*$), as defined in ref. 14, are much lower than $H_{C2}$ values, e.g., at 20 K the respective values are 4.5 T and 8.5 T. The $H_M^*$ for our sample is depicted in inset I of Fig. 5. We estimated the $J_c$ of our sample by using Bean's critical state model. Our sample studied for magnetization was in a cylindrical form, hence we used the formula $J_c = 30 \times \Delta M/d$. Where $\Delta M = |M_+| - |M_-|$ comes from the measured *M(H)* loops and *d* is the diameter of the cylindrical sample. For our sample the stabilized $\Delta M$ has a value of around 1100 emu/cm$^3$ within 1 T > *H* > -1 T at 5 K, 10 K, and 20 K. Taking *d* = 0.11 cm, the value of $J_c$ comes out to be 2.4 x 10$^5$ A/cm$^2$ which seems to be a competitive value. Note that further stabilization of flux jumps can result in much higher $J_c$ values. The plots of $J_c$ as a function of $H$ at different values of $T$ are shown in the inset II of Fig. 5. As seen from the figure, at all temperatures, $J_c$ is found to decrease with applied field. The decrease gets more rapid with increasing H and T. Further, we should mention that, the presence of flux jumps (the region marked in Fig.5) lowers the $J_C$ in our samples. Similar flux jumps at low temperatures have been earlier reported in doped $MgB_2$ compounds[16,17]. The fluxoid jumps are supposed to occur only in case of extremely high critical currents and very low heat capacity, resulting in localized motion of magnetic flux[18].

In conclusion, we have shown that bulk $MgB_2$ produced by Fe-tube encapsulated and vacuum ($10^{-5}$ torr) annealed (750 $^0$C) route has a competitive $J_c$ that is practically invariant up to 20 K within field range of 1T $>H>$ -1T. Stabilizing the flux jumps can further enhance the $J_c$ of these samples. The irreversibility field $H_M{}^*$ is found to be much smaller than $H_{c2}$. The observation of *PME* effect does point out the presence of micron or smaller normal/insulating impurities in the samples that may be responsible for high $J_c$ in the samples. However, further optimization of synthesis is required to increase the $H_M{}^*$.

Authors from NPL appreciate the interest and advice of Prof. Vikram Kumar (Director) NPL in the present work. The work is partly supported by INSA-JSPS bilateral exchange program.

**FIGURE CAPTIONS**

Fig. 1 X-ray diffraction pattern for 750 $^0$C annealed $MgB_2$ compound.

Fig. 2 Resistivity versus temperature $\rho$ (*T*) plots under various applied fields. The inset shows the extended $\rho$ (*T*) plots of the same.

Fig. 3 d$\rho$/d*T* versus *T* plots under various applied field. The inset shows the $H_{c2}(T)$ plot.

Fig. 4 Dc magnetic susceptibility versus temperature $\chi(T)$ plot in both zero-field-cooled (ZFC) and field-cooled (FC) situations at H=5 Oe. The inset shows the same for FC situation with H=1000 Oe.

Fig. 5 Magnetic hysteresis *M(H)* loops plots at 5, 10, 20 and 30 K with applied fields (*H*) of up to ± 7 Tesla. Inset-I shows the expanded *M(H)* loops to mark the $H_M^*$. Insets – II shows the $J_c$ (*H*) plots.

Fig. 1 Awana et al.

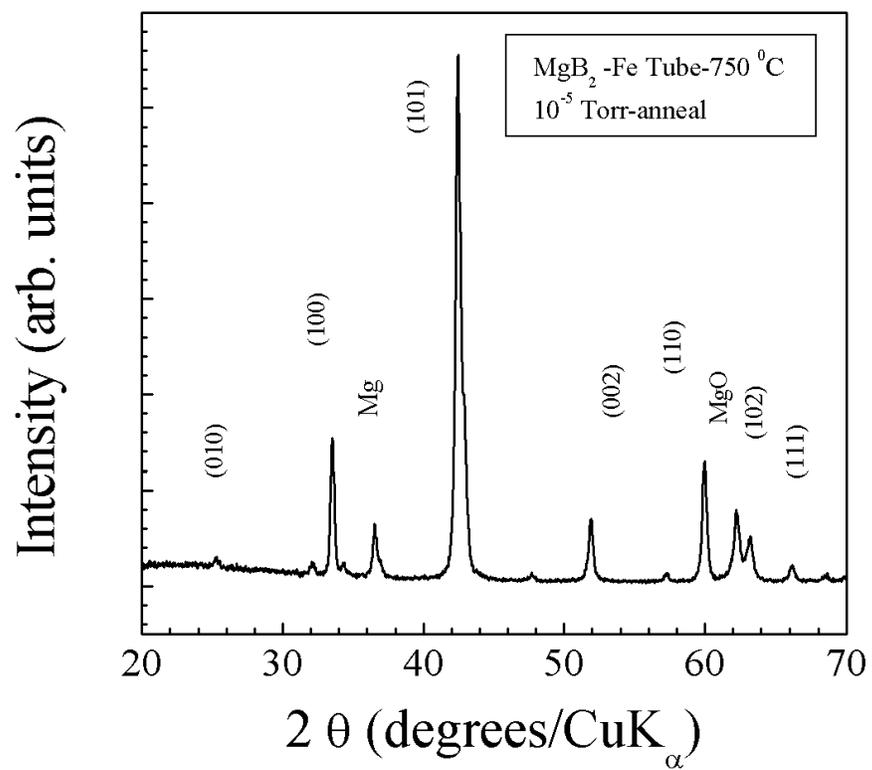

Fig. 2 Awana et al.

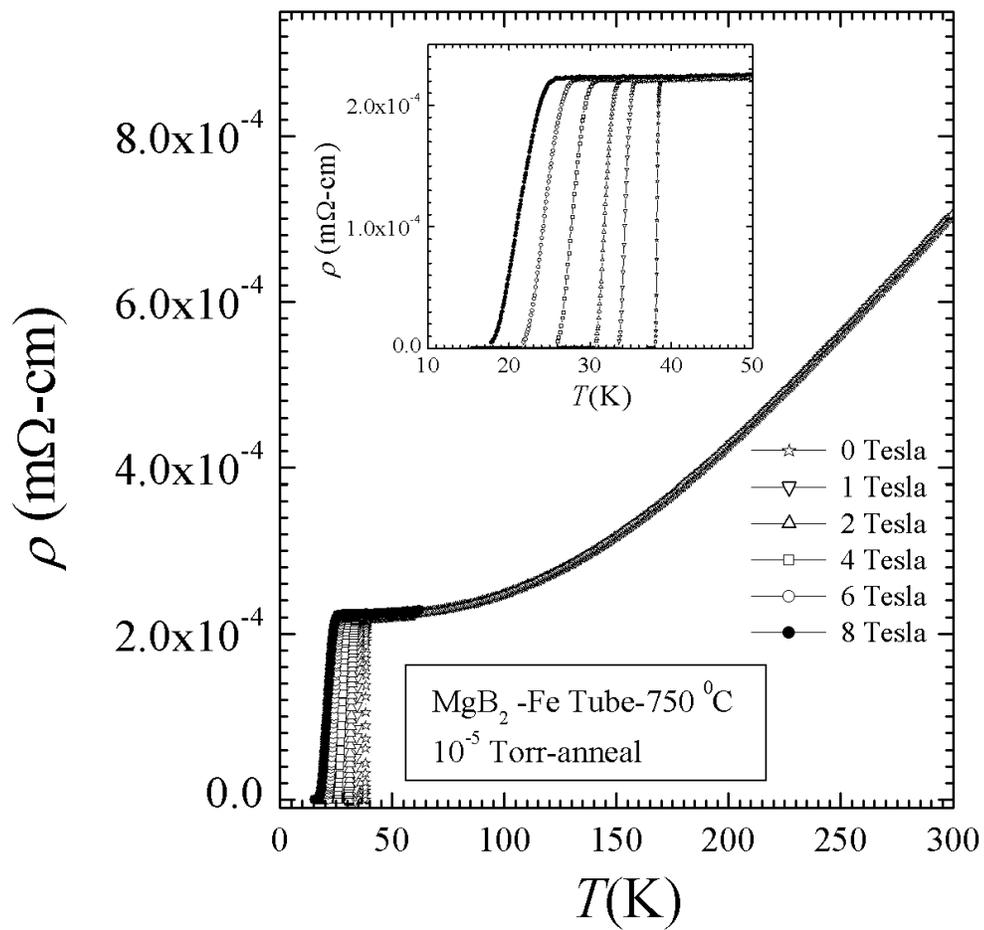

Fig. 3 Awana et al

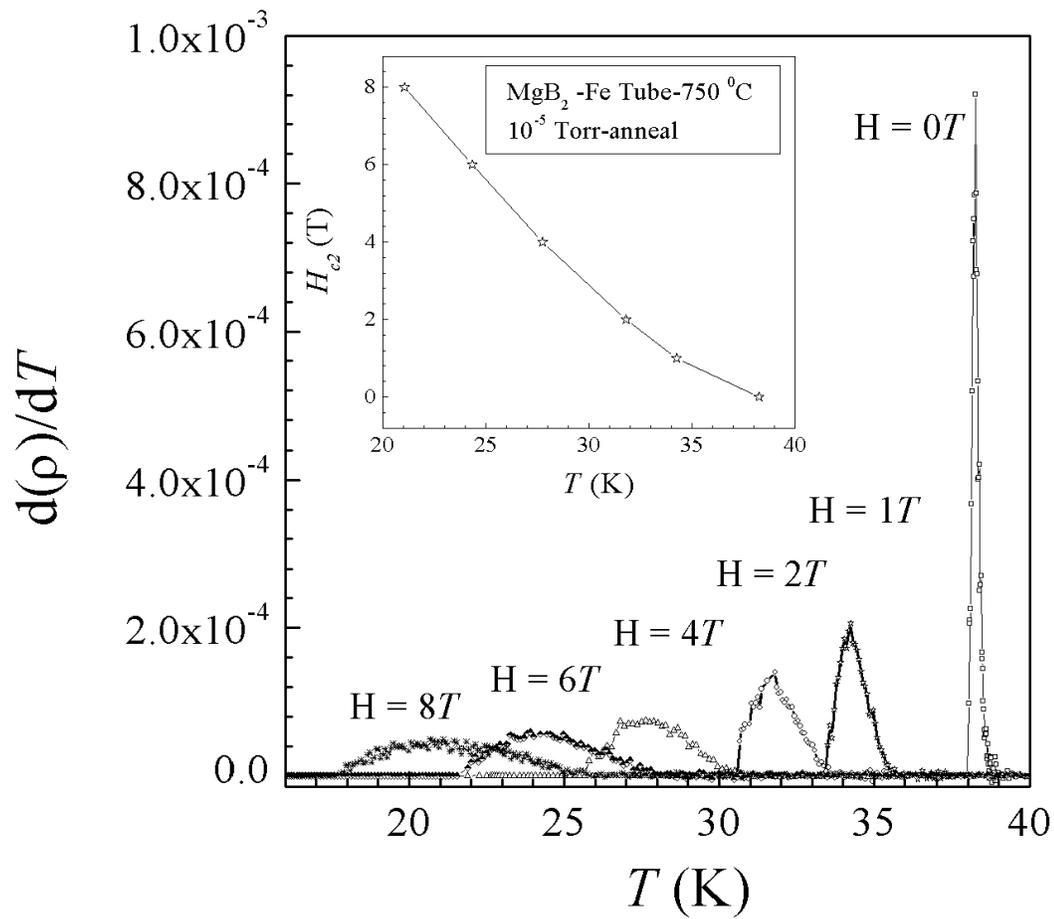

Fig. 4 Awana et al.

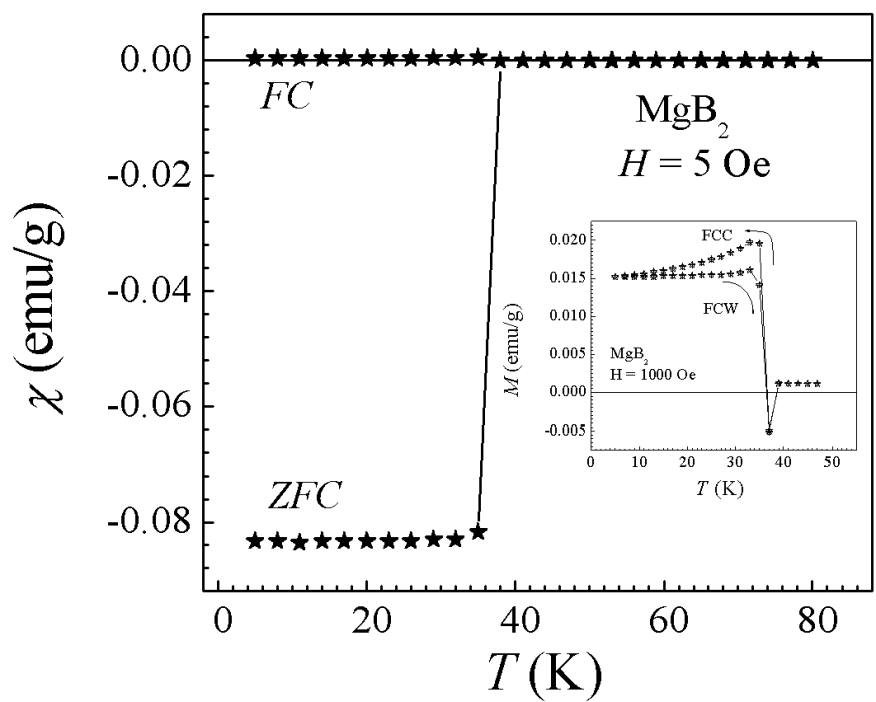

Fig. 5 Awana et al.

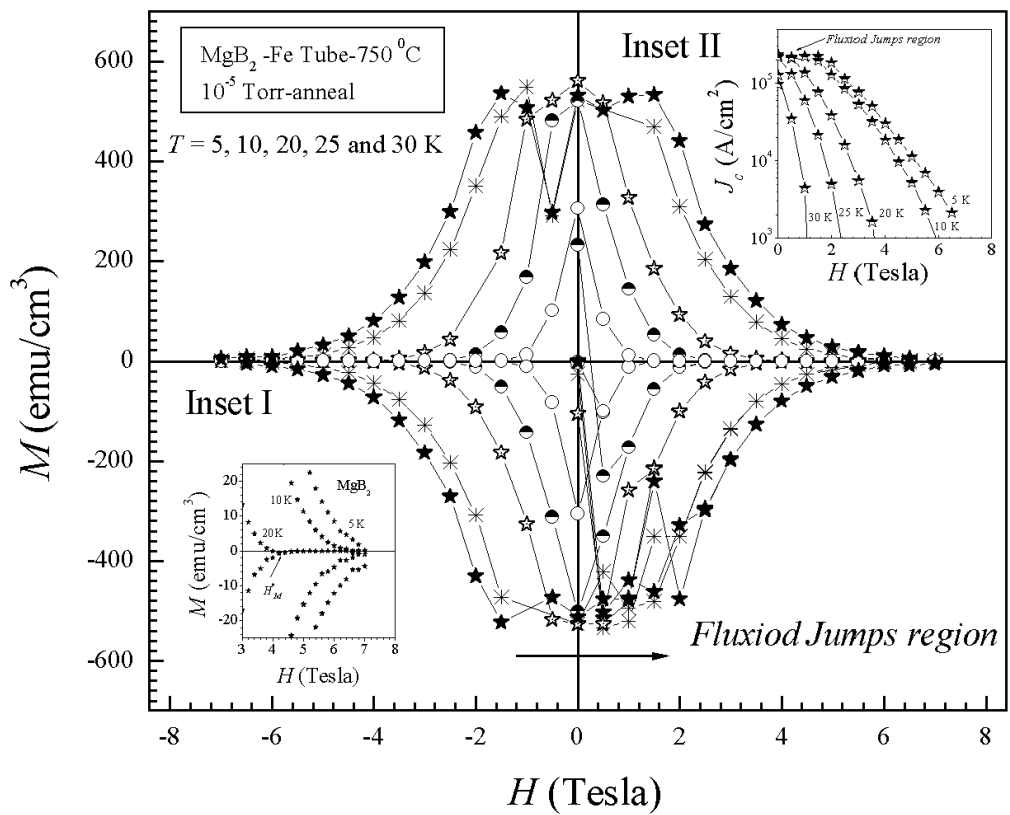

13